\documentclass[12pt]{extarticle}

\usepackage{preamble}
\usepackage{packages}

\definecolor{darkgray}{rgb}{0.33, 0.33, 0.33}

\title{\Large Accessing Large Global Charge via the $\epsilon$-Expansion}
\author[1]{Masataka Watanabe}
\affil[1]{\small Albert Einstein Center for Fundamental Physics, Institute for Theoretical Physics, University of Bern,
Sidlerstrasse 5, CH-3012 Bern, Switzerland}
\date{}

\allowdisplaybreaks[4]

\begin{document}

  \maketitle 
     \abstract{
We compute the lowest operator dimension $\Delta(J;D)$ at large global charge $J$ in the $O(2)$ Wilson-Fisher model in $D=4-\epsilon$ dimensions, to leading order in both $1/J$ and $\epsilon$.
While the effective field theory approach of \cite{Hellerman:2015nra} could only determine $\Delta(J;3)$ as a series expansion in $1/J$ up to an undetermined constant in front of each term, this time we try to determine the coefficient in front of $J^{3/2}$ in the $\epsilon$-expansion.
The final result for $\Delta(J;D)$ in the (resummed) $\epsilon$-expansion, valid when $J\gg 1/\epsilon \gg 1$, turns out to be
\begin{equation*}
\Delta(J;D)=\left[\frac{2(D-1)}{3(D-2)}\left(\frac{9(D-2)\pi}{5D}\right)^{\frac{D}{2(D-1)}}\left[\frac{5\Gamma\left(\frac{D}{2}\right)}{24\pi^2}\right]^{\frac{1}{D-1}} \epsilon^{\frac{D-2}{2(D-1)}}\right]\times J^{\frac{D}{D-1}}+O\left(J^{\frac{D-2}{D-1}}\right)
\end{equation*}
where next-to-leading order onwards were not computed here due to technical cumbersomeness, despite there are no fundamental difficulties. 
We also compare the result at $\epsilon=1$,
\begin{equation*}
\Delta(J)=0.293\times J^{3/2}+\cdots
\end{equation*}
to the actual data from the Monte-Carlo simulation in three dimensions \cite{Banerjee:2017fcx}, and the discrepancy of the coefficient $0.293$ from the numerics turned out to be $13\%$.
Additionally, we also find a crossover of $\Delta(J;D)$ from $\Delta(J)\propto J^{\frac{D}{D-1}}$ to $\Delta(J)\propto J$, at around $J\sim 1/\epsilon$, as one decreases $J$ while fixing $\epsilon$ (or vice versa), reflecting the fact that there are no interacting fixed-point at $\epsilon=0$.
Based on this behaviour, we propose an interesting double-scaling limit which fixes $\lambda\equiv J\epsilon$, suitable for probing the region of the crossover.
I will give $\Delta(J;D)$ to next-to-leading order in perturbation theory, either in $1/\lambda$ or in $\lambda$, valid when $\lambda\gg 1$ and $\lambda\ll 1$, respectively.
      }
      \newpage
\setcounter{tocdepth}{2}
\setcounter{secnumdepth}{4}
\tableofcontents
\hypersetup{linkcolor=PaleGreen4}
\newpage

\section{Introduction}
\label{sec:intro}


Not all interesting quantum field theories are solvable, nor even approximately so.
However, this does not mean we cannot perform a controlled calculation of some of their physical quantities.
It has especially been known that when the system has a global symmetry, operator dimensions of or \ac{ope} coefficients including operators of high charge can be computed to any given order in perturbation theory in the inverse charge expansion~\cite{Hellerman:2015nra,Alvarez-Gaume:2016vff,Monin:2016jmo}.\footnote{
The large-spin expansion of the light-cone bootstrap is parallel to this phenomena~\cite{Komargodski:2012ek, Fitzpatrick:2012yx,Alday:2016njk,Alday:2016jfr}, so we should try to understand them as large-quantum-number expansion as a whole.
}

This method, called {the large-charge expansion}, simply uses the idea of effective field theory.
Effective field theory can be written down by listing all operators obeying the symmetry of the system, whereby in the large-charge expansion, the large global charge $J$ (or the charge density $\rho$) gives a natural scaling of such allowed operators.
Quite remarkably, this effective theory turns out to be semi-classical and weakly-coupled as it has the large separation of scales.
For example, in $D$ dimensions, the resulting effective Lagrangian will have its \ac{uv} scale at $\Lambda_{\rm UV}\equiv {\rho}^{\frac{1}{D-1}}$ and \ac{ir} at $\Lambda_{\rm IR}\equiv 1/R_{\rm geometry}$ for theories without moduli space of vacua.
This large hierarchy of scales, $\Lambda_{\rm IR}/\Lambda_{\rm UV}\propto J^{-\frac{1}{D-1}}$, suppresses quantum corrections and higher derivative terms, and the effective action at low energies becomes classically conformally invariant.

So powerful is the classical Weyl invariance that it strongly limits the kind of operators allowed in the effective action.
For the $O(2)$ \ac{wf} fixed point at large baryon number or the $\mathbb{C}P^{N-1}$ models at large monopole number in three dimensions, it was argued in \cite{Hellerman:2015nra, Dyer:2015zha} that there are only two, one at $O(J^{3/2})$ and the other at $O(J^{1/2})$, allowed effective operators in the effective action at order $O(1)$ or above.

Now, the peculiarity of such a method using effective Lagrangian is that one cannot determine the coefficient in front of each effective operator.
They usually are some $O(1)$ numbers which cannot be computed in a controlled fashion unless there are some other weakly coupled parameters (like the large-$N$ or the $\epsilon$-expansion).
This is by no means a limitation of this methodology --
Rather, the lesson it offers is that even when the underlying theory is strongly-coupled, one can at least determine physical quantities to any given orders in $1/J$ expansion, up to some undetermined constants which depend on what theory one started with.

What is more interesting is that one can start from two different \acp{cft} and end up with the same scaling for the physical quantities, up to theory dependent constants which were left undetermined.
We could call such collection of theories, whose content of the effective action becomes the same  (and hence the $1/J$ expansion of physical quantities is the same modulo coefficients), as the large-charge universality class.
There are countably infinite many known examples of different \acp{cft} belonging to the same large-charge universality class, which are the $\mathbb{C}P^{N-1}$ models and the $O(2)$ \ac{wf} fixed point in three dimensions.

Having said that, when one is interested in a particular \ac{cft} at large global charge, one should compute those theory dependent coefficients in some way or the other.
Monte-Carlo simulations are known to work well for such purposes, which (surprisingly) showed a remarkable fit down to $J=1$, for the $O(2)$ and the $O(4)$ \ac{wf} fixed point \cite{Banerjee:2017fcx,Banerjee:2019jpw}.

When the theory in question has another weakly-coupled parameter, one can also utilise it to derive those coefficients.
Previous studies have used the $1/N$-expansion for the three dimensional $\mathbb{C}P^{N-1}$ models \cite{Dyer:2015zha, delaFuente:2018qwv}, the $\epsilon$-expansion for the three-dimensional QED at large monopole number \cite{Chester:2015wao}, the expansion in $\varepsilon\equiv N_f/N_c-11/2$ for four dimensional gauge-Yukawa models in the Veneziano limit \cite{Orlando:2019hte}, or the $1/N$-expansion for the three dimensional $O(N)$ \ac{wf} fixed point.\footnote{I thank Domenico Orlando for privately communicating me the result.}

The topic of the present paper is to proceed along this direction, to compute analytically the lowest operator dimension at leading order in $J$, for the \ac{wf} fixed-point in $D=4-\epsilon$.
This by no means is a trivial extension of the previous computations:
Using the $\epsilon$-expansion at large charge is peculiar compared with other expansions as it requires partial resuumations of the series, as argued in \cite{Chester:2015wao}.
This is because EFT (or dimensional analysis) suggests that $\Delta(J)$ scales like $\Delta(J)\propto J^{\frac{4-\epsilon}{3-\epsilon}}$, so that the $\epsilon$-expansion effectively becomes  an expansion in terms of $\epsilon \log J$.
This means unless we resum the series to all orders in the perturbation series, we are meaninglessly restricted to the region where $\frac{1}{\epsilon}\gg \log J$, which of course is not valid when analysing the $D=3$ \ac{wf} fixed point at large charge.

The surprise does not end here -- Take exactly $\epsilon=0$ in the expression $\Delta(J)=J^{\frac{4-\epsilon}{3-\epsilon}}$, and one gets $\Delta(J)\propto J^{\frac{4}{3}}$ in four dimensions, which is far from true because the theory is just a free theory in $D=4$;
One should instead get $\Delta(J)\propto J$.\footnote{This type of behaviour at large charge is typical of theories with moduli space of vacua. For more information, see \cite{Hellerman:2017veg, Hellerman:2017sur,Hellerman:2018xpi}.}
This suggests that there must be a transition of the scaling behaviour as one decreases $\epsilon$, which diverts the exponent away from $4/3$ to $1$, and we should be interested in what kind of transition this is.
I will reveal that this transition is a crossover and not a sharp phase transition of any kind, in the main body of the text.
Although this paper seems to be the first one to show there is such a crossover, but this could be potentially interesting.
For example, the fact that the there are no sharp transitions as one diverts away from free field theory means that the $\epsilon$-expansion can be useful in identifying the low-energy field content of the large-charge effective theory in terms of the field content of the original \ac{uv} theory.

I will also show that this crossover occurs when $J=\lambda/\epsilon$ (where $\lambda$ is a constant), and that the scaling gradually changes from $1$ to $\frac{4-\epsilon}{3-\epsilon}$ as one increases $\lambda$ from $0$ to $\infty$.
This suggests a double-scaling limit
\begin{equation}
\epsilon\to 0 \text{ and } J\to \infty, \qquad \text{while } \lambda\equiv \epsilon J \text{ fixed}
\end{equation}
where the lowest operator dimension behaves like
\begin{equation}
\Delta(J)\propto J^{\sigma(\lambda)}, \qquad \text{where } \sigma(0)=1 \text{ and } \sigma(\infty)=\frac{3}{2}.
\end{equation}
This double scaling limit naturally coincides the one already noticed in \cite{Libanov:1994ug, Libanov:1995gh, Son:1995wz}.

Meanwhile, I will also compute the leading order coefficient for the scaling in the $\epsilon$-expansion, when $\lambda\gg 1$.
This, when we plug in $\epsilon=1$, gives an analytical estimate on the operator dimension at large charge of the three-dimensional \ac{wf} theory.
The value turns out to be
\begin{equation}
\Delta(J;3)=0.293\times J^{3/2}+\cdots
\end{equation}
and surprisingly this is only $13\%$ away from the numerically fitted value of the coefficient, 0.337, found in \cite{Banerjee:2017fcx}.
Although this accuracy is not good enough to explain the unreasonable effectiveness (in that the fit works even at $J=1$) of the large-charge effective action, this could be a first step towards that goal.


The rest of the paper is organized as follows.
In Section \ref{2}, I will review the effective field theory at large charge of the $O(2)$ \ac{wf} fixed point in general dimensions $2<D<4$, and show that there are only two effective operators which contributes at or above $O(J^0)$.
In Section \ref{3}, I will use this information about the effective field theory to track the RG flow directly at large charge and small $\epsilon$, and compute the lowest operator dimension, to leading order in both $1/J$ and $\epsilon$.
I will find that there is a crossover in the scaling behaviour of the lowest operator dimension even at leading order, when $J$ becomes $J\sim 1/\epsilon$, and point out an interesting double-scaling limit, which fixes $J\epsilon$ to a constant.
I will also reveal that the competing size of the conformal coupling and the potential term is responsible for this crossover.
Finally, I will uncontrollably plug $\epsilon=1$ to the expression of $\Delta(J;D)$ and compute the lowest operator dimension of the $D=3$ \ac{wf} fixed point at large charge, which was then compared with the previous result from the Monte-Carlo simulation.

{\bf Note:} Three days before this paper was submitted, another paper similar in spirit appeared \cite{Arias-Tamargo:2019xld}, which dealt with the similar double scaling limit but with $\epsilon J^2$ fixed.
I also learned of a paper in preparation, which should appear on the same day as mine, by Badel, Cuomo, Monin, and Rattazzi which seems to be largely overlapping what I did here.

\section{Large-charge effective action of the $O(2)$ model in $2<D<4$}
\label{2}

\subsection{The field content of the effective field theory at large charge}

We consider, as in \cite{Hellerman:2015nra}, the $O(2)$ \ac{wf} fixed-point, but this time in general dimensions $2<D<4$, which is known to exist as an interacting fixed point \cite{Wilson:1973jj}.
The existence of interacting fixed points now makes it possible to apply directly the method of large-charge expansion for the \ac{wf} fixed point in dimensions other than three.
Below we quickly review the construction of the effective action at large charge for the \ac{wf} fixed point in general dimension.
For further information or references, consult the original paper \cite{Hellerman:2015nra}.

Let us start from the complex $\phi^4$ theory in the \ac{uv}:
\begin{equation}
\mathcal{L}_{\rm UV}=-\abs{\partial\phi}^2+{g}|\phi^4|
\end{equation}
As always, giving large charge to the system is equivalent to setting a large dimensionful \ac{vev} to the radial field $a\equiv\abs{\phi}$.
This in turn gives large mass to the $a$-field, so that it should be integrated out at large charge.
The remaining field is the angular field $\chi\sim\chi+2\pi$, which is the massless Goldstone mode from the symmetry breaking induced by the aforementioned \ac{vev}.
The \ac{vev} for the Goldstone mode is $\chi_0\equiv\Braket{\chi}=\omega t$, where $\omega$ is the chemical potential which fixes the charge density, and is proportional to the induced mass of the $a$-field.

What is important is that in the deep \ac{ir}, the effective action should not only be conformally invariant, but also be classically conformally invariant.
This is because the \ac{ir} dynamics is free and quantum corrections at scale $\Lambda$ only comes in positive powers in $\Lambda/\Lambda_{\rm UV}$, where $\Lambda_{\rm UV}\propto \omega$, which is large compared to the \ac{ir} scale.
This, combined with the fact that the $\chi$-field has dimension $0$, constrains the form of operators appearing in the effective action a lot.
Especially, at leading order in the $J$-scaling, the effective action should be simply
\begin{equation}
\mathcal{L}=b_{\chi}\abs{\partial \chi}^{D}+\cdots,
\end{equation}
where $b_{\chi}$ is an undetermined constant, as in the case of the three-dimensional \ac{wf} fixed point.
As was emphasized in the introduction, this constant should be analytically computable once we have a weak-coupling parameter, which is exactly what we will do in later sections.

\subsection{Sorting effective operators at large charge}

\heading{Rules for sorting operators}

Because of the fact that the effective action must be classically Weyl-invariant, we have the following rules for allowed operators.
\begin{itemize}
\item The term must have Weyl weight $D$.
\item The term must be $O(2)$ invariant (\it i.e., \rm it must respect the shift symmetry of $\chi$).
\item The term must be charge-conjugation invariant, \it i.e., \rm invariant under $\chi\leftrightarrow-\chi$.
\item Only $|\partial\chi|$ can appear in the denominator, because it is the mass for the $a$-field.
\end{itemize}

The semi-classical leading order action tells us the scaling of operators in terms of the charge density $\rho$, which we now fix and take large (in units of the size of the underlying geometry).
The rules to keep track of in counting the $\rho$-scaling are the following.
\begin{itemize}
\item $\partial\chi\propto \rho^{\frac{1}{D-1}}$
\item $\partial\cdots\partial\chi\propto \rho^{-\frac{D}{2(D-1)}}$
\item The leading order equation of motion, $\partial_\mu\left(|\partial\chi|^{D-2}\partial^\mu\chi\right)=0$, can be used.
\end{itemize}
The second rule comes from the $\rho$-scaling of the fluctuation of $\chi$, as the \ac{vev} for $\chi$ vanishes upon acting on more than one derivatives.
The last rule is because whenever such a combination appears, it can be replaced by something of the lower $\rho$-scaling. 

One thing one should potentially be careful about is the meaning of operator listing in fractional dimensions.
In this paper, I employ the hypothesis that one only allows fractional powers of $\abs{\partial\chi}$, as one does not seem to be able to generate fractional powers of anything else, from the original Lagrangian.
One might even be able to check this statement in the $\epsilon$-expansion. 
Also, assuming that the analytic continuation in $D$ has nice properties in the limit $D\to\infty$, excluding behaviours including trigonometric functions in $D$, one sees that fractional powers of anything other than $\abs{\partial\phi}$ must be excluded.
In other words, such a hypothesis for fractional $D$ is sufficient to reproduce the scaling rules for any integer $D$.

\heading{Effective operators at large charge in general dimensions}

We are going to list operators that are bigger than or equal to $O(J^0)$ at large charge.
First notice that we can only schematically allow for operators of the form
\begin{equation}
\frac{\partial^{n}\left[\left(\partial\chi\right)^m\right]}{\abs{\partial\chi}^{n+m-D}},
\label{eq:degreecounting}
\end{equation}
aside from terms including the curvature.
The $\rho$-scaling of the operator of this form is
\begin{equation}
\Delta\equiv\frac{(2-\ell)D-2n}{2(D-1)},
\end{equation}
where $\ell$ indicates how many $\partial\cdots\partial\chi$ there are in the numerator, and $1\leqslant\ell\leqslant \min \left(n,m\right)$ when $n\geq 1$ ($\ell$ can only be $0$ when $n=0$, trivially).

Because we are looking for operators that does not vanish in the large charge limit, we impose $\Delta\geqslant 0$, or
\begin{equation}
n\leqslant \frac{(2-\ell)D}{2}
\end{equation}
This brings down the number of operators to consider to finite, and we can examine them one by one.
In Table \ref{aaa} we show the table for the values of allowed $(n,\ell)$ and the resulting $\rho$-scaling (we assume $D\leqslant 4$ in the list).
\begin{table}[htbp]
\begin{center}
  \begin{tabular}{|c||c|c|c|c|} \hline
    $(n,\ell)$ & $\rho$-scaling & $D=4$ & $D=3$ & $D=2$ \\ \hline \hline
    $(0,0)$ & $\displaystyle \frac{D}{D-1}$ & $\displaystyle \frac{4}{3}$ & $\displaystyle \frac{3}{2}$ & $\displaystyle 2$\\ \hline
    $(1,1)$ & $\displaystyle \frac{D-2}{2(D-1)}$ & $\displaystyle \frac{1}{3}$ & $\displaystyle \frac{1}{4}$ & $\displaystyle 0$\\ \hline
    $(2,1)$ & $\displaystyle \frac{D-4}{2(D-1)}$ & $\displaystyle 0$ & $\displaystyle -\frac{1}{4}$ & $\displaystyle -1$\\ \hline
  \end{tabular}
  \caption{We show the $\rho$-scaling of operators of the form \eqref{eq:degreecounting} which do not vanish in the large-charge limit, assuming $D\leqslant 4$.}
  \label{aaa}
\end{center}
\end{table}

We are now ready to sort operators according to the $\rho$-scaling.
\paragraph{Order $\rho^{\frac{D}{D-1}}$}
The only operator at this order is 
\begin{equation}
\abs{\partial\chi}^{D}
\end{equation}
which is the leading order contribution.
\paragraph{Order $\rho^{\frac{D-2}{D-1}}$}
The only operator at this order (on the non-warped geometry) is 
\begin{equation}
{\tt Ric}_3\abs{\partial\chi}^{D-2}
\end{equation}
which, to be precise, must be Weyl-completed by $\abs{\partial\chi}^{D-4}\left(\partial\abs{\partial\chi}\right)^2$, but this has a scaling that is lower than $O(\rho^0)$ and we discard it.
\paragraph{Order $\rho^{\frac{D-2}{2(D-1)}}$}
There are two operators at this order, but they do not appear in the effective Lagrangian because they are odd under the charge conjugation symmetry.
\paragraph{Order $\rho^{\frac{D-4}{2(D-1)}}$}
There are superficially two operators at this order, but they are the form of the total derivative modulo operators smaller than $O(\rho^0)$, so there are no operators at this order.
\paragraph{Order $\rho^{0}$}
Especially, there are no operators at this order, when we assume $2\leqslant D\leqslant 4$.

To sum up, the operator sorting goes exactly the same as in the case of $D=3$, except that one multiplies everything with $\abs{\partial\chi}^{D-3}$.

\subsection{The lowest operator dimension at large charge in $D$ dimensions}

Using the above result for the effective operators, the lowest operator dimension at large charge goes as follows:
\begin{equation}
\Delta(J)=c_{\rm leading}J^{\frac{D}{D-1}}+c_{\rm Ricci}J^{\frac{D-2}{D-1}}+\gamma(D)+\cdots,
\end{equation}
where $\gamma(D)$ is the universal one-loop Casimir energy at order $O(J^0)$.
In three dimensions, this is known to take a value $\gamma(D=3)=-0.094\dots$ \cite{Hellerman:2015nra,Monin:2016bwf}.

\section{The $\epsilon$-expansion of the lowest operator dimension at large charge}
\label{3}

\subsection{Simplification of the RG flow at large charge}

If one is only interested in the leading order result in the $\epsilon$-expansion, the computation of the lowest operator dimension at large charge is most easily done by directly tracking the flow of the renormalization group.
The argument can be thought of as the precise version of what was given in Sec. 2 of \cite{Hellerman:2015nra}.
An important remark is that all statements below should be only understood at leading order in the $\epsilon$ expansion.

Let us start from the renormalised Lagrangian as follows:
\begin{equation}
\mathcal{L}=Z_{\phi}\abs{\partial\phi}^2+m^2Z_{m^2}\abs{\phi}^2+\frac{\mu^{\epsilon}gZ_{g}}{6}\abs{\phi}^4,
\end{equation}
and we use dimensional regularization and minimal subtraction in regularizing and renormalising the divergences.
As one starts out from \ac{uv}, the value for the coupling constant $g$ quickly reaches its attractive fixed point, $g_{\star}$, which is already known in the $\epsilon$-expansion as \begin{equation}
g_{\star}=\frac{24\pi^2}{5}\epsilon.
\end{equation}
One can also consider fine-tuning the coupling so that we already are on the fixed point in the \ac{uv}.

Now, turning on the \ac{vev} for the field $a\equiv\abs{\phi}$ changes the renormalization group flow when $\mu\sim m_a$, after which the RG evolution of $\mu^{\epsilon} g$ completely freezes.
Here, $m_a$ is the mass of the $a$-field induced by the \ac{vev}, which is $(m_a)^2\propto\mu^\epsilon g\abs{a}^2$
This process is usually not under control, and one cannot usually see what value $\mu^\epsilon g$ takes.
However, in the $\epsilon$ expansion, because such a shift of $g_{\star}$ due to the \ac{vev} starts only at order $O(g_{\star}^2)=O(\epsilon^2)$, one can just plug in the value of $g^{\star}$ into $\mu^{\epsilon}g$.
As one still do not know what value of $\mu\sim O(m_a)$ one should plug in, we will just plug in $\mu=K\times m_a=K\times \mu^{\epsilon/2}g_{\star}^{1/2}\abs{a}$, where $K$ is some $O(1)$ constant.
Solving for $\mu$, we get
\begin{equation}
\mu^\epsilon =\left(K\sqrt{g_{\star}}\right)^{\frac{\epsilon}{1-\epsilon/2}}a^{\frac{\epsilon}{1-\epsilon/2}},
\end{equation}
and therefore we generate the potential that goes as
\begin{equation}1
V(a)=\left(K\sqrt{g_{\star}}\right)^{\frac{\epsilon}{1-\epsilon/2}}\times \frac{g_{\star}}{6}\times a^{2+\frac{2}{1-\epsilon/2}}=\frac{g_\star}{6}\abs{\phi}^{4+\frac{\epsilon}{1-\epsilon/2}}\times\left(1+O(\epsilon)\right)
\end{equation}

\subsection{Semi-classical computation}

\heading{The semi-classical Lagrangian}

At large charge, the Lagrangian including this potential term can be treated semi-classically, which at leading order in $O(\epsilon)$ reads, on the unit sphere,
\begin{equation}
\mathcal{L}=\abs{\partial\phi}^2+\frac{(D-2)^2}{4}\abs{\phi}^2+\frac{g_\star}{6}\abs{\phi}^{\frac{2D}{D-2}}
\end{equation}
where the second term is the conformal coupling of the scalar field.
It is also very important to \it not \rm expand $\abs{\phi}^{\frac{2D}{D-2}}$ in terms of $\epsilon$.
It would give terms like $(\epsilon \log a)^n$ but from effective field theory analysis, we know they must finally resum to $a^{\frac{2D}{D-2}}$ in the end.
In other words, we conduct computation assuming $\epsilon$ is not small, except that we use the $O(\epsilon)$ result for the coefficient in front of $\abs{\phi}^{\frac{2D}{D-2}}$, as this value is contaminated at $O(\epsilon^2)$ by $\epsilon \log K$ or the running of the coupling at around $\mu\sim m_a$, in the presence of the \ac{vev}.

Now we solve the equation of motion at fixed \ac{vev} for the $a$-field, assuming the helical configuration for the lowest energy state, $\phi=a\times e^{i\omega t}$.
The equation of motion then gives
\begin{equation}
\omega^2=\frac{g_\star D}{6(D-2)}a^{\frac{4}{D-2}}+\frac{(D-2)^2}{4}
\end{equation}

The charge $J$ and the energy on the unit sphere $\Delta$ goes as follows, in terms of $a$.
\begin{eqnarray}
J&=&\alpha(D)\sqrt{\frac{2g_\star D}{3(D-2)}}a^{\frac{2(D-1)}{D-2}}\times \sqrt{1+\frac{3(D-2)^3}{2D}\frac{1}{g_\star a^{\frac{4}{D-2}}}}\label{eq:J-a}\\
\Delta&=&\frac{\alpha(D)g_\star}{3}\frac{(D-1)}{D-2}a^{\frac{2D}{D-2}}\times \left(1+\frac{3(D-2)^3}{2(D-1)}\frac{1}{g_\star a^{\frac{4}{D-2}}}\right),\label{eq:E-a}
\end{eqnarray}
where $\alpha(D)\equiv \frac{2\pi^{D/2}}{\Gamma(D/2)}$ is the area of the unit $(D-1)$-sphere.
This is sufficient to infer the relation between $\Delta$ and $J$.
Because it is analytically hard to compute this directly, we will take $g_\star a^2$ large or small and compute $\Delta(J)$ in the form of the Taylor expansion in terms of it.
The scaling structure from such an analysis will be 
\begin{equation}
\Delta(J)=c_{\rm leading}J^{\frac{D}{D-1}}+c_{\rm Ricci}J^{\frac{D-2}{D-1}}+\cdots
\end{equation}
when $g_\star a^2 \gg 1$ (which is what happens in $D=3$), and
\begin{equation}
\Delta(J)=J^{1}+c_{\rm Ricci}J^{0}+\cdots
\end{equation}
when $g_\star a^2 \ll 1$.

\heading{The effect of the conformal coupling and the crossover}

Because there are two competing small parameters in the system, $g_\star\sim\epsilon$ and $1/a$, the relative size of those scales becomes important.
We examine below the cases where $\epsilon a^\frac{4}{D-2}$ scales as $\epsilon a^\frac{4}{D-2}=\lambda a^p$, seperately when $0<p<\frac{4}{D-2}$, $p=0$, and $p<0$.

\paragraph{(a) $p>0$: Semi-classical regime at $J\gg 1/\epsilon$}

In this case, because we take $a$ large, we can approximate \eqref{eq:J-a} and \eqref{eq:E-a} as follows:
\begin{eqnarray}
J_{p>0}&=&\alpha(D)\sqrt{\frac{2g_\star D}{3(D-2)}}a^{\frac{2(D-1)}{D-2}}\\
\Delta_{p<0}&=&\frac{\alpha(D)g_\star}{3}\frac{(D-1)}{D-2}a^{\frac{2D}{D-2}},
\end{eqnarray}
The relationship between $\Delta$ and $J$ therefore becomes
\begin{eqnarray}
\Delta(J)&=&c_{0}(D)\times {J}^{\frac{D}{D-1}}\\
c_{0}(D)&=&\frac{2(D-1)}{3(D-2)}\left(\frac{3(D-2)}{8D\pi}\right)^{\frac{D}{2(D-1)}}\Gamma\left(\frac{D}{2}\right)^{\frac{1}{D-1}}g_\star^{\frac{D-2}{2(D-1)}},
\end{eqnarray}
and the condition $\epsilon a^\frac{4}{D-2}=\lambda a^p$ can be rewritten as
\begin{equation}
J\propto \frac{1}{\epsilon^{\frac{1}{2}\left(\frac{6}{2-p}-1\right)}}\gg \frac{1}{\epsilon}
\quad (p>0)
\end{equation}

\paragraph{(b) $p=0$: Crossover region at $J\sim 1/\epsilon$}

One should in theory be able to express $\Delta$ in terms of $J$, but it would be too cumbersome.
We can understand this region $J\propto \frac{1}{\epsilon}$ as the crossover region from region (a), $J\gg \frac{1}{\epsilon}$, to region (c), $J\ll \frac{1}{\epsilon}$.
The analysis of this regime will be done in perturbation theory in $J\epsilon\ll 1$ and in  $(J\epsilon)^{-1}\ll 1$ in the double-scaling limit section.
 
\paragraph{(c) $p<0$: Free theory regime at $J\ll 1/\epsilon$}

In this region, what dominates takes over, and we can approximate \eqref{eq:J-a} and \eqref{eq:E-a} as follows:
\begin{eqnarray}
J_{p<0}&=&\alpha(D)(D-2)a^2\\
\Delta_{p<0}&=&\frac{\alpha(D)(D-2)^2}{2}a^2,
\end{eqnarray}
so that
\begin{equation}
\Delta(J)=\frac{D-2}{2}J,
\end{equation}
whose coefficient of course is the mass dimension for the scalar field in $D$ dimensions.
The condition $\epsilon a^\frac{4}{D-2}=\lambda a^p$ translates to
\begin{equation}
J\propto \frac{1}{\epsilon^{\frac{2}{2+\abs{p}}}}\ll \frac{1}{\epsilon} \quad (p<0)
\end{equation}

To summarise, $\Delta(J)$ behaves in the following way, depending on how large $J$ is:
\begin{eqnarray}
\Delta(J)=
\begin{cases}
J^\frac{D}{D-1} & \text{semi-classical behaviour at }J\gg \frac{1}{\epsilon}\\
J^{\sigma(J)} & \text{crossover at } J\sim \frac{1}{\epsilon}\\
J^1 &\text{free theory behaviour at } J\ll \frac{1}{\epsilon}
\end{cases}
\end{eqnarray}

\subsection{The double-scaling limit}

One can use various double-scaling limit for this system at large $J$ and small $\epsilon$, depending on what regime one is interested in.
For example, when one is interested in the regime (a), one can use the double-scaling limit with $\epsilon J^{1+\abs{\alpha}}$ fixed.
This will ensure that the limit taken leads to the operator scaling 
\begin{equation}
\Delta\propto J^{\frac{4-\epsilon}{3-\epsilon}}+\cdots \quad (\text{when }J^{1+\abs{\alpha}} \text{ is fixed})
\end{equation}
On the other hand, one can use the double-scaling limit like $\epsilon J^{1-\abs{\alpha}}$ fixed, to reach the region (c).
This will in turn  ensure that the limit taken leads to the operator scaling 
\begin{equation}
\Delta\propto J^1 +\cdots \quad (\text{when }J^{1-\abs{\alpha}} \text{ is fixed})
\end{equation}

Somewhat more interesting is the double-scaling limit which fixes $\lambda\equiv \epsilon J$, which accesses the crossover region (b), in the regime of weak-coupling.
The operator scaling will take the form
\begin{equation}
\Delta(J)\propto J^{\sigma(\lambda)}+\cdots,
\end{equation}
where $\sigma(\lambda)$ is an increasing function in $\lambda$, with $\sigma(0)=1$ and $\sigma(\infty)=\frac{D}{D-1}$.\footnote{I thank Zohar Komargodski for making me notice that presenting this as a double-scaling limit is interesting.
}

\heading{Perturbative expansion in the double-scaling limit}

We treat cases where $\lambda\gg 1$ and $\lambda\ll g$ separately, and see what are the correction to the leading formula in $1/J$ and in $\epsilon$ in the double-scaling limit.

\paragraph{When $J=\frac{\lambda}{\epsilon}$ and $\lambda\gg 1$}

By using the Taylor expansion, we get
\begin{equation}
\Delta(J;D)=c_0(D=4)\times J^{4/3}\left(1+\left(\frac{5\sqrt{2}}{3\sqrt{3}}\cdot \frac{1}{\epsilon J}\right)^{2/3}+\left(\frac{5}{6\sqrt{3}}\cdot \frac{1}{\epsilon J}\right)^{4/3}+\cdots\right)
\end{equation}

\paragraph{When $J=\frac{\lambda}{\epsilon}$ and $\lambda\gg 1$}

By using the Taylor expansion, we get
\begin{equation}
\Delta(J;D)=J^{1}\left(1+\frac{\epsilon J}{10}-\frac{\left(\epsilon J\right)^2}{50}+\cdots\right)
\end{equation}

\subsection{Three-dimensional Wilson-Fisher fixed point at large charge}

Let us now plug $\epsilon=1$ and see what leading coefficient one finds in front of the $J^{3/2}$ dependence of the lowest operator dimension.
We can, as in \cite{Hellerman:2015nra}, ignore the conformal coupling, which is equivalent to analysing the semi-classical region, $J\gg 1/\epsilon=1$.
Note that fractional powers inside the expression of $c_{0}(D)$ will not be truncated to $O(\epsilon)$ but considered to be exact, which is the expectation from the EFT at large charge.


Using the expression for $c_0(D=3)$ and plugging in $g_\star =\left.\frac{24\pi^2\epsilon}{5}\right|_{\epsilon=1}=\frac{24\pi^2}{5}$, we have
\begin{equation}
c_0(D)=\frac{1}{3^{3/4}5^{1/4}}.
\end{equation}
This gives the lowest operator dimension at leading order at large charge as
\begin{equation}
\Delta(J)=0.293\times J^{3/2}+\cdots
\end{equation}
whose value is different from the result of the Monte-Carlo simulation found in \cite{Banerjee:2017fcx}, which is $\Delta(J;D)=0.337 J^{3/2}$, by $13\%$.

\heading{Comparison with the $\mathbb{C}P^{0}$ model}

Because the \ac{ir} fixed point of the $\mathbb{C}P^{0}$ model is the $O(2)$ \ac{wf} fixed point, we can use the result from the large-$N$ expansion of the $\mathbb{C}P^{N-1}$ model \cite{delaFuente:2018qwv} and again just plug in $N=1$.
The result yields
\begin{equation}
\Delta(J)\propto 0.312\times J^{3/2},
\end{equation}
which gives a slightly better number compared to ours.

\section{Conclusions and Outlook}

I have computed the lowest operator dimension at large charge $J$ in $D=4-\epsilon$ dimensional \ac{wf} theory to leading order in both $1/J$ and $\epsilon$, with corrections that go as powers of $(J\epsilon)^{-1}\ll 1$ or $J\epsilon\ll 1$.
Especially, by extrapolating to $D=3$ the result of the computation, I computed, including numerical coefficient the lowest operator dimension of the $D=3$ \ac{wf} model to leading order in the $1/J$-expansion:
\begin{equation}
\Delta(J)=0.293\times J^{3/2}+\cdots.
\end{equation}
I also compared the result with the previous Monte-Carlo result and found a discrepancy of $13\%$.

I also found an interesting crossover in the scaling behaviour of $\Delta(J)$, and found that it scales $\Delta(J)\propto J^{\frac{D}{D-1}}$ when $J\gg 1/\epsilon$ and $\Delta(J)\propto J^1$ when $J\ll 1/\epsilon$. 
I also pointed out that the crossover can be accessed in the weak-coupling limit by taking a double-scaling limit, where $1/\epsilon,\, J\to \infty$ while $\epsilon J$ fixed.

There are a number of important future directions.
\begin{description}
\item[Non-trivial check of the method of large-charge expansion]\mbox{}\\
The lowest operator dimension expanded in $1/J$ in the large-charge universality class of the $O(2)$ \ac{wf} fixed point has a theory-independent part at order $O(J^0)$.
Going to higher-orders in $\epsilon$ to reproduce this number gives a consistency check of the method of the large charge expansion.
Computing and comparing the subleading coefficients to the numerical data will be of great importance too.
\item[Checking dualities at large charge]\mbox{}\\
If two theories are dual to each other, the lowest operator dimension at charge $J$ should match to all orders in $1/J$-expansion, including the coefficients.
Computing the coefficients will give more precise check of the duality than just looking at the scaling behaviour.
\item[Contributions from massive modes]\mbox{}\\
Massive modes (in this example, the $a$-field), contributes as $O(\exp\left(-m_a\right))$, where $m_a\propto 1/J^{\alpha}$.
This can be explicitly seen by actually computing the free energy as a function of the chemical potential $\mu$ for the charge density -- one can directly see the exponential contribution, coming from the one-loop energy shift of the massive $a$-field. 
One should also compare the result with the double-scaling limit (which fixes the mass of the BPS dyon) in rank-$1$ SCFTs introduced in \cite{Bourget:2018obm}.
This will be reported in the forthcoming paper.
\end{description}

\section*{Acknowledgements}
\begin{sloppypar}
The author is grateful to Andrew Gasbarro, Simeon Hellerman, Nozomu Kobayashi, Zohar Komargodski, Keita Nii, Domenico Orlando and Susanne Reffert for valuable discussions.
The author also thank the Simons Center for Geometry and Physics for hospitality during the conference ``Quantum-Mechanical Systems at Large Quantum Number'' while this paper was being completed.
\end{sloppypar}

\bibliographystyle{JHEP} 
\bibliography{references,cond-mat}

\end{document}